# The specific heat and the radial thermal expansion of bundles of single-walled carbon nanotubes

**M.I. Bagatskii, M.S. Barabashko, A.V. Dolbin, V.V. Sumarokov**

*B. Verkin Institute for Low Temperature Physics and Engineering of the National Academy of Sciences of Ukraine*

*47 Lenin Ave., Kharkov 61103, Ukraine*

**and B. Sundqvist**

*Department of Physics, Umea University, Umea SE - 901 87, Sweden*

E-mail: bagatskii@ilt.kharkov.ua

The specific heat at constant pressure $C(T)$ of bundles of single-walled carbon nanotubes (SWNTs) closed at their ends has been investigated in a temperature interval of 2 – 120 K. It is found that the curve $C(T)$ has features near 5 K, 36 K, 80 K and 100 K. The experimental results on the $C(T)$ and the radial thermal expansion coefficient $\alpha_R(T)$ of bundles of SWNTs oriented perpendicular to the sample axis have been compared. It is found that the curves $C(T)$ and $\alpha_R(T)$ exhibit a similar temperature behavior at T>10 K. The temperature dependence of the Gruneisen coefficient $\gamma(T)$ has been calculated. The curve $\gamma(T)$ also has a feature near 36 K. Above 36 K the Gruneisen coefficient is practically independent of temperature ($\gamma \approx 4$). Below 36 K $\gamma(T)$ decreases monotonically with lowering temperature and becomes negative at T< 6 K.

## 1. Introduction

Single-walled carbon nanotubes (SWNTs) are convenient objects on which to investigate the dynamics of phonons in low-dimensional systems. As theoretical studies of the thermal properties of SWNTs [1-10] show, the temperature behavior of the heat capacity of these objects is much affected by the chirality and diameters of tubes and their number within a bundle. Isolated SWNTs have a one-dimensional structure with four acoustic modes (one longitudinal, two transverse and one torsional) obeying a linear dispersion law. At low temperatures the phonon part of the heat capacity of three – dimensional solids has a cubic $T$ – dependence. In isolated SWNTs the degrees of freedom corresponding to the azimuth vibrations of the carbon atoms are frozen at low temperatures and the phonons belong to a strictly one – dimensional system. As a result, the heat capacity has a linear dependence on temperature [1].

The low-energy part of the spectrum of isolated SWNTs still remains practically unstudied experimentally. While being prepared, SWNTs experience Van der Waals attractive forces and form three – dimensional anisotropic bodies – SWNT bundles [10]. We can therefore expect a close-to-cubic temperature dependence for the phonon part of the heat capacity of SWNT bundles in the low temperature region. As the temperature increases and the higher energy levels are successively occupied, the behavior of the phonon heat capacity changes from quasi – three – dimensional to quasi – two – dimensional.

The low temperature heat capacity of SWNT bundles was measured in Refs. [10 – 14]. The starting powders were prepared by laser vaporization (LV) [15] (samples in [10-14]) and arc-discharge (AD) [16] (samples in [10, 14]), respectively. In accordance with Refs. [15, 16], the samples [10- 14] contained 70 % - 90 % of SWNT bundles, other carbon forms and catalysts [17 -19]. In Refs. [10 -14] the heat capacity was measured by the thermal relaxation technique and the calorimeter was cooled using $^4$He as a heat exchange gas. The presence of even a small amount of $^4$He in the calorimeter cell can degrade the accuracy of measurement at helium temperatures [11, 14]. The discrepancy between data in Refs. [10-14] is equal to about 40 % above 20 K and even to several times at helium temperatures.

The goal of this study was to investigate the low temperature dynamics of SWNT bundles by calorimetric methods In the future we are planning to investigate the dynamics of various adsorbed gases in SWNT bundles. The heat capacity data taken on pure SWNT bundles will be helpful in analyzing the results of the experiments planned.

## 2. Experiment

The specific heat at constant pressure ($P \cong 0$) of SWNT bundles closed at their ends has been investigated in the temperature interval 2-120 K using an adiabatic calorimeter [20, 21].

A cylindrical sample (7.2 mm high, 10 mm in diameter, of 1.2 g/cm$^3$ density) was prepared by compressing SWNT plates under the pressure 1.1 GPa. The plates (~0.4 mm thick) were obtained by compacting a SWNT powder ("Cheap Tubes") under P=1.1 GPa. It is expected that the sample prepared by this technique has a pronounced anisotropy of certain properties in the directions perpendicular and parallel to the sample axis. The pressure 1.1 GPa applied to a 0.4 mm SWNT plate aligns the nanotubes in the plane normal to the pressure vector, the average deviation being ~4 degrees [22].

The thermal expansion along the sample axis is determined only by the radial component. The technique of sample preparation has no effect on the results of the heat capacity measurement because these pressures cause no change in the structure of SWNT bundles.

The powder was prepared by chemical catalytic vapor deposition (CVD). It contained over 90 wt % of SWNT bundles, other allotropic forms of carbon (fullerite, multiwalled nanotubes and amorphous carbon) and about 0.6 % of cobalt catalyst [23]. The average tube diameter in the sample was 1.1 nm, the average length of the SWNT bundles was 15 μm [24]. The number of nanotubes in the bundles varied within 100 – 150 (estimated from high – resolution TEM pictures).

Previously, this sample was used in dilatometric measurements of the radial thermal expansion coefficient [25]. After the measurement of α_R(T), the sample was placed in a hermetized vessel. Prior to this calorimetric investigation the sample was washed several times with pure $N_2$ gas and held in dynamic vacuum at $T\sim500$ K for ~ 12 hours. After cooling the sample to room temperature, the vessel was filled with pure $N_2$ gas under the pressure P ~ 1.1 atm. The mass of the sample was found by analytical balance weighing to be m= 716.0±0.05 mg.

The sample weighing and mounting in the calorimeter along with hermetization of the vacuum chamber of the calorimeter took about three hours. The vacuum chamber of the calorimeter was then washed several times with pure $N_2$ gas and the sample was held in the dynamic vacuum at T=355 K for 10 hours. The calorimeter was cooled through wires without using $^4$He as an exchange gas. After cooling the calorimeter to T=120K, the cryostat bath with the calorimeter – insert in it [20, 21] was filled with liquid helium. Cooling the calorimeter to helium temperatures took about five hours.

The calorimetric experiment was controlled with a personal computer (PC) using a data management and acquisition system based on a Keithley 2700/7700 multimeter with an Advantech PCI – 1671 interface board communicating through an IEEE – 488 (GPIB) data bus.

The temperature of the calorimeter was measured with a calibrated CERNOX CX – 1010 – SD – 0.3 D resistance thermometer (Lake Shore Cryotronics). The characteristic variation of the calorimeter temperature during the time of heating was $\Delta T \approx 0.05\ T$, where $T$ is the temperature of the calorimeter before heating.

The thermal contact between the sample and the calorimetric vessel was improved using Apiezon – N grease. First we measured the heat capacity of the calorimetric vessel with the Apiezon – N grease but without a sample ("empty calorimeter"). The heat capacity of the sample of SWNT bundles was found by subtracting the heat capacity of the "empty" calorimeter from the heat capacity of the calorimeter with the sample.

The contribution of the sample to the total heat capacity of the calorimeter with the sample was 75% at 2 K, 60 % at 4 K, 40 % at 10 K and about 30 % at $T$=20-120 K. The error in the measurement of the heat capacity Cs of the sample of SWNT bundles was 18 % at $T$=2 K, 2% at $T$=5 K, and decreased to 1 % as the temperature increased to 120 K. The specific heat C($T$) of the SWNT bundles was calculated using the expression C($T$)/m.

We estimated the difference between the specific heats at constant pressure ($C_P$) and constant volume ($C_V$) of SWNT bundles at $T$=120 K using the expression $\Delta C = C_P - C_V = \beta^2 VT/\chi_T$, the molar volume V = 9*10$^{-6}$ m$^3$/mol [26] and the isothermal compressibility $\chi_T$ =2.7*10$^{-11}$ Pa$^{-1}$ [27]. Since data on the coefficient of volumetric thermal expansion $\beta$ of SWNT bundles oriented perpendicular to the sample axis are unavailable, we assumed that $\beta \approx 3\alpha_R = 9*10^{-5}$ K$^{-1}$, where $\alpha_R$ is the coefficient of radial thermal expansion [25]. Thus $\Delta C$ is approximately equal to 1.6% of the heat capacity of the sample.

**3. Results and discussion**

The experimental temperature dependence of the specific heat C(*T*) of SWNT bundles closed at their ends taken in the interval 2 – 120 K is illustrated in Fig. 1. It is seen that the curve increases monotonically with rising temperature in the whole temperature interval. For comparison, literature data [10 – 14] are added. In the temperature interval 30 – 120 K our results are systematically 20 – 28 % higher than the data of Ref. [12] and 10 – 18 % lower than the data of Ref. [10]. Below 15 K the heat capacities given in Refs. [10-14] are systematically higher than ours. The variation of the heat capacity magnitudes can be brought about by both experimental errors and different techniques of sample preparation. Depending on the preparation method, samples can have different compositions, different contents of nanotubes per bundle, different tube distributions in diameter and chirality. The information about measurement techniques, measurement temperatures, methods of powder preparation, characteristics of nanotubes and SWNT bundles, sample mass and composition is summed up in the table. The heat capacity of SWNT bundles depends on the diameter and chirality of the nanotubes and on the average number in a bundle [1, 6 – 9]. The difference among the heat capacity magnitudes due to these factors increases at low temperatures.

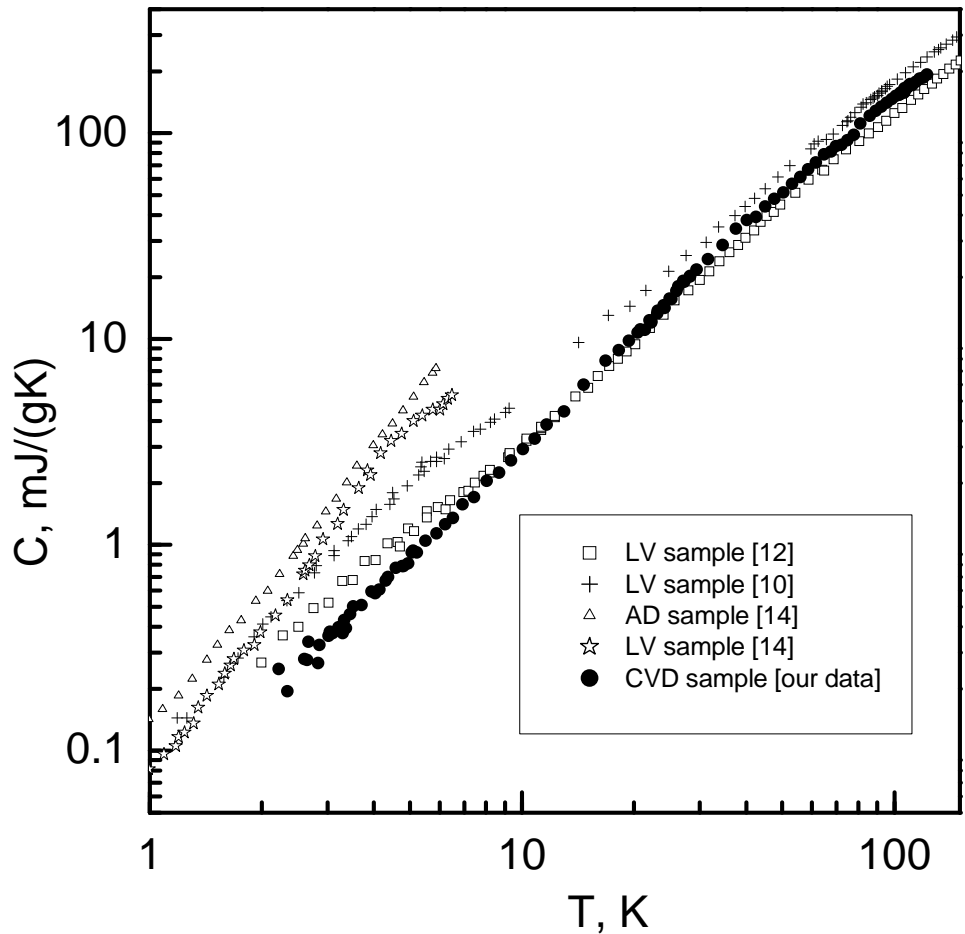

Fig. 1 Measured heat capacities of SWNT bundles from this work and Refs. [10], [12] and [14].

Table. Measurement techniques, measurement temperatures, powder preparation methods, characteristics of SWNTs and SWNT bundles, sample masses and compositions.

| Samples | This work | Hone [12] | Mizel [10] | Lasjaunias [14] | |
|---|---|---|---|---|---|
| Manufacturer | «Cheap Tubes», USA | «tubes@rice», USA | Rice University, USA | Rice University, USA | Montpellier, France |
| Techniques of synthesis | Catalytic chemical vapor deposition (CVD Method) | Laser vaporization technique (LV) | Laser vaporization technique (LV) | Laser vaporization technique (LV) | Arc discharge technique (AD) |
| Average tube diameter | 1.1 nm | 1.25 nm | 1.3 nm | 1.4 nm | 1.4 nm |
| Amount of nanotubes | Over 90 wt.% | – | 70–90 wt.% | 70–90 wt.% | 70–90 wt.% |
| Catalyst fraction | Co(0.60 at.. %) | Co/Ni(2 at. %) | Co/Ni (1.2 at. %) | Co/Ni (2 at %) | Ni (0.5 at %) Y (0.5 at %) |
| Average quantity of nanotubes in a bundle | 100-150 | 40-50 | 100-500 | 30-100 | 20-30 |
| Technique of measuring heat capacity | Adiabatic technique | Thermal relaxation technique | Thermal relaxation technique | Thermal relaxation technique | Thermal relaxation technique |
| Sample mass | 716.0±0.05 mg | 9.5 mg and 2.5 mg | 10-20 mg | 45 mg | 90 mg |
| Temperature interval | 2-120 K | 2-300 K | 1-200 K | 0.1-6 K | 0.1-6 K |

Depending on their chirality index (n, m) [28-31], single-walled carbon nanotubes can be either metallic or semiconducting..As follows from theoretical calculations [1], at low temperatures the contribution of phonons to the heat capacity of SWNTs with metallic conduction exceeds that of electrons by several orders of magnitude.

The SWNT bundles used in our experiment were prepared by chemical vapor deposition (CVD) and most of the tubes had metallic conduction [34]. According to studies [35], CVD – prepared samples contain about 60 % of "armchair" (n, n) tubes having metallic conduction [30], about 15% of "zigzag" (m, 0) tubes [30] (one third of metallic tubes and two – thirds of semiconducting tubes) and "chiral" (m, n) nanotubes. It is shown [34] that the CVD – prepared powder used for producing our sample contains nanotubes with (m, n) chirality for which the difference m-n is a multiple of 3, implying metallic conductivity.

The chirality parameters (m, n) and the nanotube diameter D are related as [30]

$$D = \sqrt{m^2 + n^2 + mn} \cdot \sqrt{3} d_0 / \pi \tag{1}$$

where $d_0 = 0.142$ nm is the distance between the neighboring C atoms in the graphite plane. Assuming that in our sample the average diameter of the armchair (n, n) nanotubes is D=1.1 nm, their chirality was calculated by Eq. (1) to be (8, 8).

One of the samples in Ref. [14] contained SWNT bundles prepared by the arc discharge technique. Such bundles have a lower concentration of "zigzag" (m, 0) nanotubes in comparison to CVD – prepared SWNT bundles (our sample) [35]. It is shown theoretically [9] that below 20 K the heat capacity of an ensemble of isolated "zigzag" (m, 0) nanotubes is lower than that of an ensemble of isolated "armchair" (n, n) nanotubes. This may explain partially why our heat capacities are lower than the results of [14].

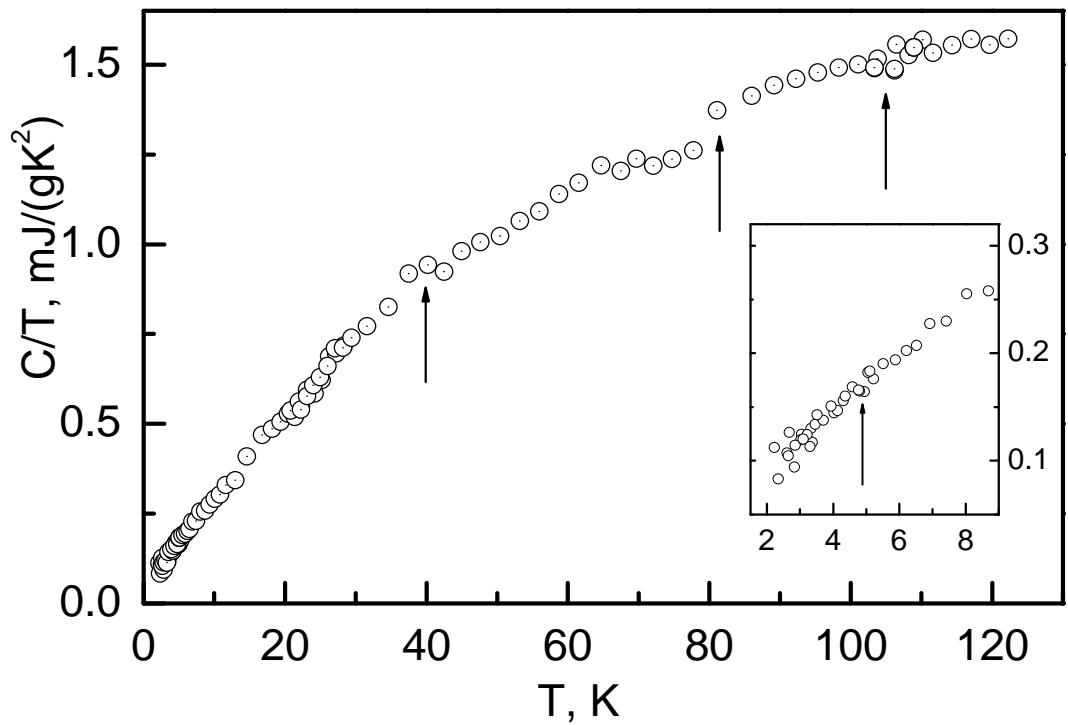

Fig. 2. The temperature dependence of the heat capacity of SWNT bundles closed at their ends (our data) plotted as C/*T* vs. *T*.

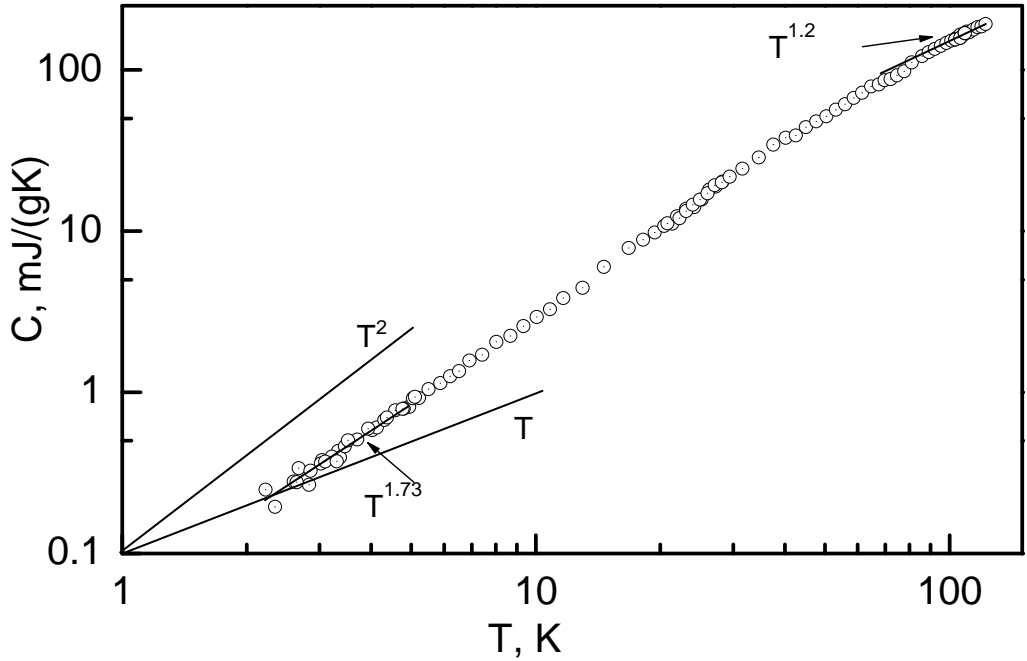

Fig. 3. The temperature dependence of heat capacity of SWNT bundles (our data) plotted as log C(*T*) vs. log *T*.

Our results on the heat capacity of SWNT bundles are shown plotted as C(*T*)/*T* vs. *T* in Fig. 2. The inset in Fig. 2 shows the low temperature region on an expanded scale. It is seen that the curve C(T)/T has features (breaks) near 5 K, 36 K, 80 K and 100 K (shown by arrows). The curve has a shallow depression in the interval 60 – 80 K. The same data are shown plotted as log C(*T*) vs. log *T* in Fig. 3. It is seen that in the region 2 K≤*T*≤5 K the experimental curve C(T) shows a temperature dependence C(T) ~ $T^{1.73}$. This behavior is characteristic of quasi-two-dimensional systems. In the interval 108≤ *T*≤120 K the curve C(*T*) follows the dependence C(*T*) ~ $T^{1.2}$, which is closer to that of a quasi-one-dimensional systems. A near-quasi-one-dimensional behavior of the heat capacity of SWNT bundles above 120 K was also observed in [10]. We can therefore expect a cubic temperature dependence of the heat capacity of SWNT bundles at lower temperatures. Note that the experimental heat capacity of SWNT bundles prepared by laser evaporation and arc discharge in Ref. [14] are described by the dependence C(*T*)~$T^3$ in the intervals – 0.35-4.5 K and 0.2 -1.8 K, respectively. However, the dependence C(*T*)~$T^3$ does not hold for the results of Refs. [10-12] and this study.

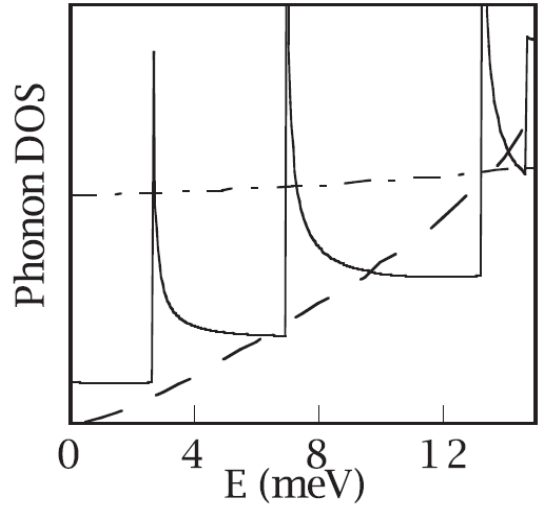

Fig. 4. Calculated phonon densities of states of graphene ( dash-dot line) [36], graphite (dashed line) [38] and an ensemble of isolated (10, 10) SWNTs (solid curve) [11,36].

The phonon contribution to the heat capacity is determined by the phonon density of states (PDOS). The theoretical PDOS curves for graphite [36], graphene [38] and an ensemble of isolated (10, 10) SWNTs [11,36] are illustrated in Fig. 4. The PDOS for graphite and graphene were also calculated in Refs. [37,39]. As follows from theoretical calculations [11, 38], the 1$^{st}$ and 2$^{nd}$ optical subbands appear in the PDOS curve $g_{SW}(E)$ of an isolated (10, 10) SWNT near $E_{1\,sub} \approx 2.7$ meV (30 K) and $E_{2\,sub} \approx 7$ meV (81 K), respectively. At E<2.7 meV $g_{SW}(E)$ is independent of E (see Fig. 4). Therefore, at low temperatures the heat capacity of an ensemble of isolated SWNTs is proportional to temperature (see Fig. 5). In this case only acoustic modes are populated. As the temperature rises the 1$^{st}$ optical subband $g_{SW}(30\,K)$ is being occupied and the character of the dependence C(T) changes from one – dimensional to quasi –two – dimensional. According to theory [1, 2], in this case the crossover temperature $T_c$ increases as the SWNT diameter decreases. $T_c$ can be estimated as $T_c \approx E_{1\,sub}/6k_B$, where $k_B$ is the Boltzmann constant [8,11].

In Fig. 5 our experimental results are compared with C(T) calculated for graphite [12], graphene [12] and an ensemble of isolated (5, 5) and (10, 10) SWNTs. In the low temperature region the curves C(T) for SWNT bundles and for ensembles of isolated SWNTs are intermediate between the corresponding curves for graphene and graphite. As the chirality changes from (10, 10) to (5, 5), the crossover temperature $T_c$ increases from ≈3 K to ≈ 7 K. Below 20 K the curve C(T) of SWNTs with chirality (5, 5) runs systematically higher than the corresponding curve for SWNTs with chirality (10, 10). Below 20 K the difference between the calculated heat capacity for ensembles of isolated (5, 5) and (10, 10) SWNTs [6] and the experimental data for bundles of SWNTs monotonically increases with decreasing temperature.

The differences in magnitude and behavior between the experimental C(T) of SWNT bundles and the theoretical $C_{SW}(T)$ of an ensemble of isolated (10, 10) SWNTs below 20 K (Fig. 5) suggest that the formation of bundles changes qualitatively the low frequency region $g_{SW}(E)$ of isolated SWNTs. The nanotubes start interacting, which affects the vibrational modes in the SWNTs and generates low frequency intertube vibration in the bundle [40-42].

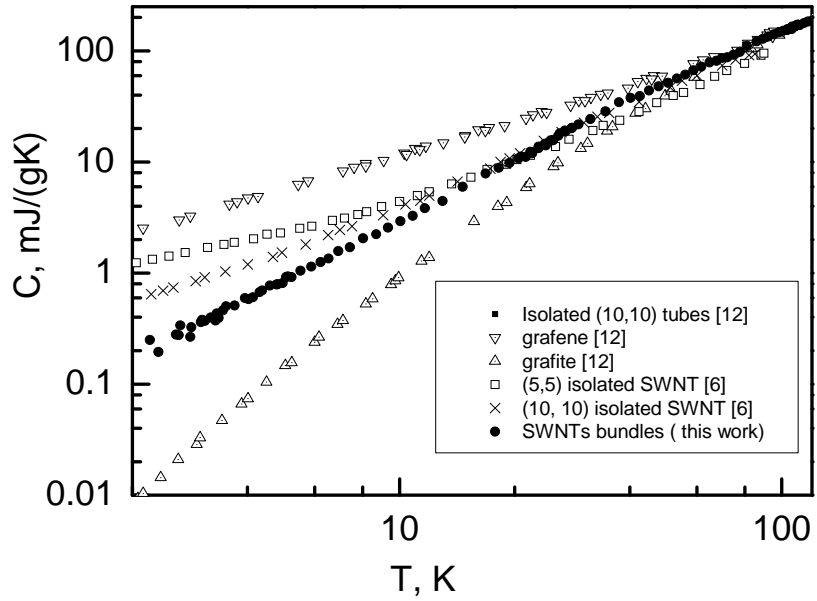

Fig. 5. Comparison of experimental (our results) and calculated heat capacities of graphite, graphene and an ensemble of isolated (5, 5) and (10, 10) SWNTs.

The dependence $g(E)$ of LV and AD prepared SWNT bundles has been investigated theoretically and experimentally (inelastic neutron scattering) in a wide interval of energies from 0.3 to 225 meV [40-42]. Below 15 meV the experiment was carried out at room temperature [41]. The $g(E)$ and $g_{SW}(E)$ profiles are similar above 15 meV and differ qualitatively below 15 meV. The unusual energy dependence $g(E)$ below 10 meV is due to the contributions of the intertube modes in the 2D triangular lattice of SWNT bundles and the intratube excitation coupling between the nearest tubes. The experiments showed that in the region 0.3 – 1.5 meV the bundles of all investigated types of SWNTs exhibited a linear dependence $g(E) \sim E$ (see Ref. [42], Fig. 20). In this case a quadratic temperature dependence $C(T) \sim T^2$ can be expected below 4 K. In the interval T=2-5 K our $C(T)$ data are described by a near – quadratic function $C(T) \sim T^{1.73}$. The feature in the curve C/T(T) at ~36 K accounts for the frequency dependence of the density of states up to 22 meV (see Ref. [42], Fig. 18).

As in the case of both the heat capacity and the thermal expansion [25], the temperature dependence of the electric conductivity [43] has an anomaly near 30 K. It is attributed [43] to a change of the conductivity mechanism as the system of oriented SWNT bundles transforms from a Luttinger liquid to a dielectric Mott phase at T=25 -36 K. Note that both the heat capacity (this study), the coefficient of the radial thermal expansion $\alpha_R(T)$ [25] and the electric conductivity [43] were measured on identical samples of SWNT bundles.

The coefficient of the radial thermal expansion $\alpha_R(T)$ of bundles of SWNTs oriented perpendicular to the sample axis [25] was investigated in the interval 2-120 K using precise low temperature dilatometry [44]. It is found experimentally [25] that $\alpha_R(T)$ is negative below 5.5 K. This behavior of $\alpha_R(T)$ can be attributed to the contribution of low frequency transverse vibrations to thermal expansion. The contribution is negative for layered structures [45,46] and nanotubes [47].The experimental curves C(T) and α_R(T) are compared in Fig. 6. The scale of α_R(T) on the ordinate was chosen so that we could match the curves C(T) and α_R(T) above 36 K. It is seen that the temperature behavior of C(T) and α_R(T) is similar but near 36 K the slope of the curve α_R(T) increases.

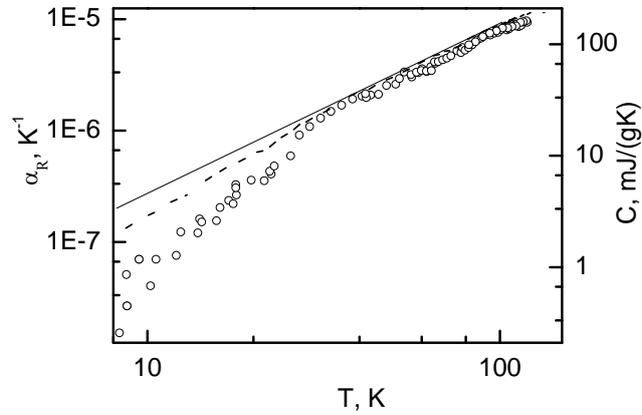

Fig. 6. Comparison of the experimental curves *C(T)* (dashed line) and *α_R(T)* (circles) [25] of bundles of SWNTs oriented perpendicular to the sample axis. The straight line is an extrapolation of the high temperature behaviour of C(T) to the region below 36 K.

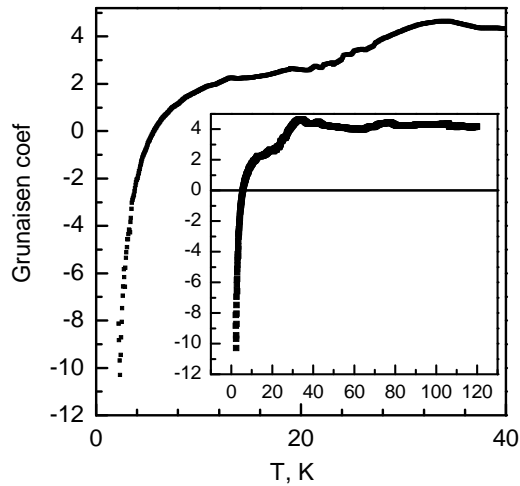

Fig. 7. The temperature dependence of the Gruneisen coefficient *γ(T)* of bundles of SWNTs oriented perpendicular to the sample axis.

The temperature dependence of the Gruneisen coefficient *γ(T)* in the interval 2 – 40 K is shown in Fig. 7. The *γ(T)* was calculated using the expression $\gamma=\beta V/\chi_T C$, assuming that $\beta \approx 3\alpha_R$ (see above the section "Experiment"), our results on *C(T)* and literature data for the thermal expansion [25], the molar volume [26] and the isothermal compressibility [27] of SWNTs. As is the case with C(T) and α_R(T), the curve has a feature near 36 K. Above 36 K γ is practically independent of temperature and is γ~4, which is somewhat higher than γ≅ 2-3 for cryocrystalls [48,49]. Below 36 K *γ(T)* decreases smoothly with lowering temperature and becomes negative at T<6 K. Below 6 K the thermal properties of SWNT bundles are determined mainly by acoustic and libration lattice modes, whereas above 36 K the contribution of optical intratube modes is dominant. The experimental curve $\gamma(T)$ agrees qualitatively with theoretical calculation [47].

## 4. Conclusions

The specific heat at constant pressure, C(T), of bundles of single-walled carbon nanotubes (SWNTs) oriented perpendicular to the sample axis has been investigated in the temperature interval 2 – 120 K. It is shown that in the intervals 2-5 K and 108 – 120 K the behavior of the heat capacity is close to that in quasi – two-dimensional and quasi-one-dimensional systems, respectively.

The heat capacity C(T) and the radial thermal expansion coefficient $\alpha_R(T)$ [25] of bundles of SWNTs oriented perpendicular to the sample axis measured in the interval 2-120 K have been analyzed. It is found that the curves C(T) and $\alpha_R(T)$ demonstrate a similar temperature behavior.

The temperature dependence of the Gruneisen coefficient $\gamma(T)$ has been calculated. As with C(T) and $\alpha_R(T)$, the $\gamma(T)$ has a feature near 36 K. The Gruneisen coefficient is practically independent of temperature above 36 K ($\gamma \cong 4$) and decreases monotonically with lowering temperature below 36 K. At T<6 K $\gamma(T)$ becomes negative.

**Acknowledgments**

The authors are grateful to Profs. V. G. Manzhelii, B. A. Danilchenko, E. S. Syrkin and S. B. Feodosyev for helpful discussions. Thanks are also due to V. B. Eselson and S. I. Popov for their assistance in the experiment.